\documentclass[12pt]{article}

\begin{document}

\title{Super-energy and Killing-Yano tensors}

\author{Ovidiu Tintareanu-Mircea \thanks{E-mail:~~~ovidiu@venus.nipne.ro} and
Florian Catalin Popa \thanks{E-mail:~~~catalin@venus.nipne.ro}\\
{\small \it Institute of Space Sciences}\\
{\small \it Atomistilor 409, RO 077125, P.O.Box M.G.-23, Magurele, Bucharest, Romania}}

\date{\today}

\maketitle

\begin{abstract}
In this paper we investigate a class of basic super-energy
tensors, namely those constructed from Killing-Yano tensors, and
give a generalization of super-energy tensors for cases when we
start not with a single tensor, but with a pair of tensors.

~


~

Pacs: 04.20.Cv

\end{abstract}

\section{Introduction}
The Bel-Robinson tensor \cite{LBell:1958, LBell:1959} is a
valuable tool in many mathematical formalisms involving the
gravitational field, such as hyperbolic formulation
\cite{HFriedrich:1996, OAReula:1998} of the Einstein field
equations \cite{AAnderson:1997, MAGBonilla:1998}, the causal
propagation of gravity \cite{AAnderson:1997,
BerqvistSenovilla:preprint, BonillaSenovilla:1997}, the existence
of global solutions of the Cauchy problem \cite{ChoquetYork:1980,
HawkingEllis:1973, ARendall:1998} and the study of the global
stability of spacetimes \cite{ARendall:1998, Christodoulou:1993}.
It was introduced for the first time to solve the lack of a
well-posed definition of a local energy-momentum for the
gravitational field, since the Principle of Equivalence tells us
that it is allays possible to choose a reference system along any
timelike curve such that the gravitational field in this
particular coordinate system vanishes and so does the
gravitational energy density. The analogy between many of it's
properties and those of energy-momentum tensor of electromagnetic
fields has led many authors to look for similar super-energy
tensors of fields other than the gravitational one
\cite{Chevreton:1964, PenroseRindler:1986, Senovilla:2000,
Teyssandier:2001}.

Killing-Yano tensors, introduced by Yano \cite{KYano:1952},
are geometrical objects closely related with {\em hidden}
symmetries \cite{GibbonsRietdijk} in pseudoclassical models
for spinning particle. They have an important role in generating
solutions for generalized St\"{a}ckel-Killing equations which in
turn lead to new constants of motion
\cite{PenroseFloyd:PhD, MVisinescuDVaman:PhysRev:1996}. Concrete applications for
Euclidean Taub-NUT space, pp-wave and Siklos metric can be found in
\cite{MVisinescuDVaman:PhysRev:1998, MVisinescu:JFA:2000, BaleanuCodoban, DBaleanu:2002}.
For the geodesic motions in
the Taub-NUT space, the conserved vector analogous to the
Runge-Lenz vector of the Kepler type problem is quadratic in
4-velocities, it's components are St\"{a}ckel-Killing tensors and
they can be expressed as symmetrized products of Killing-Yano
tensors
\cite{GibbonsRuback:1987, vanHolten:1995, MVisinescuDVaman:PhysRev:1998, MVisinescuDVaman:ForPhys}.
A particular class of Killing-Yano tensors, namely those covariantly constant which
realize certain square roots of the metric tensor, are involved in generating
Dirac-type operator and the generator of a one-parameter Lie group connecting
this operator with the standard Dirac one
\cite{Carter:1979McLenaghan:1979,MVisinescuICotaescu:ClassWuantumGrav:2004}.
An investigation of spaces admitting Killing-Yano tensors based
on covariant Petrov classification can be found in
\cite{Dietz:1981, Dietz:1982}.

The paper is organized as follows:
In Sec.\ref{preliminaries} is presented a short description
of how the super-energy tensors can be constructed starting from an arbitrary tensor.
In Sec.\ref{setens-for-ky-sect} we investigate a particular class of super-energy tensors, namely
those obtained from Killing-Yano tensors and write the Killing-Yano equations
in terms of electric and magnetic parts.
In Sec.\ref{setensor-pair-sec}
we give an extension of super-energy tensors obtained not from a single tensor, but
from a pair of tensors and present how this generalization is involved in construction
of Runge-Lenz vector for Euclidean Taub-NUT space.
Conclusions are given in Sec.\ref{conclusions-sec}.
\section{Preliminaries}\label{preliminaries}
In this section we briefly describe how, starting with an
arbitrarily tensor field, an super-energy tensor can be
constructed. An extensive and detailed presentation can be found
in \cite{Senovilla:2000}. Be $t_{\mu_1\dots\mu_m}$ an arbitrary
rank $m$ covariant tensor.

Consider the set of indices which are antisymmetric with $\mu_1$,
including $\mu_1$. This set will constitute a block of $n_1\leq n$
antisymmetric indices which will be denoted by $[n_1]$. Take the
next index of $t_{\mu_1\dots\mu_m}$ which is not in $[n_1]$ and
make a second block $[n_2]$ containing $n_2\leq n$ antisymmetric
indices. This procedure have to be continued until each index of
$t_{\mu_1\dots\mu_m}$ will belong to such a block. Thus, the $m$
indices of $t_{\mu_1\dots\mu_m}$ have been separated in $r$
blocks, each of them containing $n_{\Upsilon}$
($\Upsilon=1,\dots,r$) completely antisymmetric indices, with
$n_1+\dots +n_r =m$. This way $t_{\mu_1\dots\mu_m}$ can be seen as
a $r$-fold $(n_1,\dots,n_r)$-form which will be schematically
denoted by $t_{[n_1],\dots,[n_r]}$ where $[n_{\Upsilon}]$ specify
the $\Upsilon$-th block with $n_{\Upsilon}$ antisymmetric indices.

Having all indices of $t$ separated in $r$ disjunct blocks of totally
antisymmetric indices, a Hodge dual relative to each of these blocks can be defined as follows.
Be $t^{\{\Omega\}}{}_{\mu_1\dots\mu_p}=t^{\{\Omega\}}{}_{[\mu_1\dots\mu_p]}$
a tensor with an arbitrary number of indices without any symmetry properties denoted
schematically by $\{\Omega\}$, plus a set of $p\leq n$ completely antisymmetric indices
$\mu_1\dots\mu_p$.
The dual of $f$ relative to the $\mu_1\dots\mu_p$ indices is defined as follows
\begin{equation}\label{dual-def}
t^{\{\Omega\}}{}_{\stackrel{\star}{\mu_{p+1}\dots\mu_n}}\equiv
\frac{1}{p!}\eta_{\mu_1\dots\mu_n}t^{\{\Omega\}}{}^{\mu_1\dots\mu_p}\;,
\end{equation}
where $\star$ is placed over the group of indices on which
operation of tacking the dual acts.
With (\ref{dual-def}), the following $2^r$ duals can be defined for the
$r$-fold $(n_1,\dots,n_r)$-form $t_{[n_1]\dots[n_r]}$
\begin{equation}\label{all-duals}
t_{\stackrel{\star}{[n-n_1]}\dots[n_r]},
\dots,t_{[n_1]\dots\stackrel{\star}{[n-n_r]}},
t_{\stackrel{\star}{[n-n_1]}\stackrel{\star}{[n-n_2]}\dots[n_r]},
\dots,t_{\stackrel{\star}{[n-n_1]}\stackrel{\star}{[n-n_2]}\dots\stackrel{\star}{[n-n_r]}}\;.
\end{equation}
With the aid of these duals, a canonical {\em electric-magnetic}
decomposition of $t_{\mu_1\dots\mu_m}$ relative to any timelike unit vector field $\vec u$
can be constructed
by contracting each of the above duals with $r$ copies of $\vec u$, each one with
the first index of each block. When $\vec u$ is contracted with a {\em starred} block
$\stackrel{\star}{[n-n_\Upsilon]}$ we have a {\em magnetic part} in that block,
and an {\em electric part} when $\vec u$ is contracted with a {\em regular block}
$[n-n_\Upsilon]$.\\
For example, if we choose the dual
$t_{\stackrel{\star}{[n-n_1]}[n_2]\dots[n_r]}$,
the corresponding {\em electric-magnetic} part is denoted by
$({}_{\vec u}^tH{\underbrace{E\dots E}_{r-1}}
)_{[n-n_1-1][n_2-1]\dots[n_r-1]}$
and it's definition is
\begin{eqnarray}
&&(
{}_{\vec u}^tH{\underbrace{E\dots E}_{r-1}}
)_{\mu_{n_1}+2\dots\mu_n:\nu_2\dots\nu_{n_2}:\dots:\rho_2\dots\rho_{n_r}}
=\nonumber\\
&&\tilde t_
{\stackrel{\star}{\mu_{n_1}+1\mu_{n_1}+2\dots\mu_n}:
\nu_1\nu_2\dots\nu_{n_2}:\dots:
\rho_1\rho_2\dots\rho_{n_r}}
u^{\mu_{n_1}+1}u^{\nu_1}\dots u^{\rho_1}\;.
\end{eqnarray}
Here, $\tilde{t}_{[n_1],\dots,[n_r]}$ denotes the tensor obtained
from $t_{[n_1],\dots,[n_r]}$ by permuting it's indices such that
the first $n_1$ indices are exactly those from $[n_1]$ block, the
next $n_2$ indices are exactly those from block $[n_2]$ and so on.
The super-energy tensor associated to $t$
is defined as follows
$$
T_{\lambda_1\mu_1\dots\lambda_r\mu_r}\left\{t\right\}\equiv
\frac12
\left\{
\left(t_{[n_1]\dots[n_r]}\times t_{[n_1]\dots[n_r]}
\right)_{\lambda_1\mu_1\dots\lambda_r\mu_r}+
\right.
$$
$$
+\left(t_{\stackrel{\star}{[n-n_1]}[n_2]\dots[n_r]}\times
t_{\stackrel{\star}{[n-n_1]}[n_2]\dots[n_r]}
\right)_{\lambda_1\mu_1\dots\lambda_r\mu_r} + \dots +
$$
$$
+ \left(t_{[n_1]\dots[n_{r-1}]\stackrel{\star}{[n-n_r]}}\times
t_{[n_1]\dots[n_{r-1}]\stackrel{\star}{[n-n_r]}}\right)_{\lambda_1\mu_1\dots\lambda_r\mu_r}
+ \dots +
$$
$$
+\left(t_{\stackrel{\star}{[n-n_1]}\stackrel{\star}{[n-n_2]}[n_3]\dots[n_r]}\times
t_{\stackrel{\star}{[n-n_1]}\stackrel{\star}{[n-n_2]}[n_3]\dots[n_r]}
\right)_{\lambda_1\mu_1\dots\lambda_r\mu_r}+\dots +
$$
\begin{equation}\label{def-se-tens}
\left. +
\left(
t_{\stackrel{\star}{[n-n_1]}\dots\stackrel{\star}{[n-n_r]}}\times
t_{\stackrel{\star}{[n-n_1]}\dots
\stackrel{\star}{[n-n_r]}}
\right)_{\lambda_1\mu_1\dots\lambda_r\mu_r}\right\}\;,
\end{equation}
where
\begin{equation}\label{semi-s}
(t\times t)_{\lambda_1\mu_1\dots\lambda_r\mu_r}\equiv
\left(\prod_{\Upsilon=1}^{r}\frac{1}{(n_{\Upsilon}-1)!}\right)\,
\tilde{t}_{\lambda_1\rho_2\dots\rho_{n_1}\dots\dots
\lambda_r\sigma_2\dots\sigma_{n_r}}
\tilde{t}_{\mu_1\hspace{7mm}\dots\dots\mu_r}^{\hspace{2mm}\rho_2\dots\rho_{n_1}
\hspace{5mm}\sigma_2\dots\sigma_{n_r}}\;.
\end{equation}
\section{Super-energy tensor for Killing-Yano tensor}\label{setens-for-ky-sect}
Starting with a Killing-Yano tensor, i.e. an antisymmetric rank two tensor which satisfies
\begin{eqnarray}\label{defky}
Y_{\mu(\nu;\lambda)}&\equiv&Y_{\mu\nu;\lambda}+Y_{\mu\lambda;\nu}=0\;,
\end{eqnarray}
we obtain the following expression for the associated super-energy tensor (\ref{def-se-tens})
\begin{equation}\label{bseky}
T_{\mu\nu}\{Y_{[2]}\} = Y_{\mu a}Y_\nu^{\ a}-
\frac14g_{\mu\nu} Y_{ab}Y^{ab}\;
\end{equation}
which, taking into account the antisymmetry of $Y$, turns out to be symmetric
$T_{\mu\nu}=T_{\nu\mu}$.
Since $Tr\left(g_{\mu\nu}\right)=4$, it follows that $T_\mu^{\ \mu}\{Y_{[2]}\}=0$.
The divergence $T_{\ \mu;\nu}^{\nu}\{Y_{[2]}\}$ is not necessarily null.
Also, the super-energy tensor does not depend on dimension of the underlying space.

The terms $Y_{\mu a}Y_\nu^{\ a}$ and $Y_{ab}Y^{ab}$ lead to a St\"{a}ckel-Killing tensor
$K_{\mu\nu}\equiv Y_{\mu a}Y_\nu^{\ a}$
(a symmetric rank two tensor which satisfies $K_{(\mu\nu;\lambda)}=0$),
respective to it's trace $K_{a}^{\ a}\equiv Y_{ab}Y^{ab}$.
Thus, a super-energy tensor obtained from a Killing-Yano tensor turns out to be the
trace free part of a St\"{a}ckel-Killing tensor.
We remember that for every rank two tensor $T_{\mu\nu}$ there
is a covariant decomposition in an antisymmetric part
$T_{[\mu\nu]}\equiv\frac12\left[T_{\mu\nu}-T_{\nu\mu}\right]$,
a trace free symmetric part $T_{(\mu\nu)}-\frac14 g_{\mu\nu}T_\rho^{\ \rho}$
(where $T_{(\mu\nu)}\equiv\frac12\left[T_{\mu\nu}+T_{\nu\mu}\right]$)
and a part proportional with the metric tensor $\frac14 g_{\mu\nu}T_\rho^{\ \rho}$.

The super-energy tensor (\ref{bseky}), which we'll write
$T_{\mu\nu}\{Y_{[2]}\}=K_{\mu\nu}-\frac14g_{\mu\nu}Tr(K)$,
is not by itself a St\"{a}ckel-Killing tensor since
\begin{equation}
T_{(\mu\nu;\lambda)}\{Y_{[2]}\}=-\frac14 g_{(\mu\nu}\left(Tr(K)\right)_{,\lambda)}\;,
\end{equation}
which in general does not vanishes.

Since the super-energy tensor for $Y_{\mu\nu}$ can be written \cite{Senovilla:2000}
\begin{equation}\label{se-tens-with-dual}
T\{Y_{[2]}\}=\frac12\left[Y_{\mu a}Y_\nu^{\ a}+Y_{\;\mu a}^\star Y_{\ \nu}^{\star\;a}\right]\;,
\end{equation}
it follows that if the starting Killing-Yano tensor $Y_{\mu\nu}$ is self-dual
($Y_{\mu\nu}^\star=Y_{\mu\nu}$) or antiself-dual ($Y_{\mu\nu}^\star=-Y_{\mu\nu}$)
we have
\begin{equation}
T_{\mu\nu}\{Y_{[2]}\}=Y_{\mu a}Y_\nu^{\ a}\;.
\end{equation}
A rank $r$ tensor $Y_{\mu_1\dots\mu_r}$ is said to be a Killing-Yano tensor if it is totally
antisymmetric and satisfies
\begin{equation}
Y_{\mu_1\dots(\mu_r;\lambda)}=0\;.
\end{equation}
A rank $r$ tensor $K_{\mu_1\dots\mu_r}$ is said to be a St\"{a}ckel-Killing tensor if it is totally
symmetric and satisfies
\begin{equation}
K_{(\mu_1\dots\mu_r;\lambda)}=0\;.
\end{equation}
The following contraction
\begin{equation}
K_{\mu\nu}\equiv Y_{\mu\rho_2\dots\rho_r}Y^{\ \rho_2\dots\rho_r}_{\nu}
\end{equation}
is a rank-$2$ St\"{a}ckel-Killing tensor.\\
The super-energy tensor for the Killing-Yano tensor $Y_{\mu_1\dots\mu_r}$ is
\begin{equation}\label{se-tensor-of-gen-ky}
T_{\mu\nu}\left\{Y_{[r]}\right\}=
\frac1{(r-1)!}
\left(
Y_{\mu\rho_2\dots\rho_r}Y_{\nu}^{\ \rho_2\dots\rho_r}-
\frac1{2r}g_{\mu\nu}
Y_{\rho_1\rho_2\dots\rho_r}Y^{\rho_1\rho_2\dots\rho_r}
\right)\;.
\end{equation}
Since $Y_{\mu\rho_2\dots\rho_r}Y_{\nu}^{\ \rho_2\dots\rho_r}$
and $Y_{\rho_1\rho_2\dots\rho_r}Y^{\rho_1\rho_2\dots\rho_r}$
are a St\"{a}ckel-Killing tensor $K$, respective it's trace, it follows that
(\ref{se-tensor-of-gen-ky}) is the trace free part of $K$.

We saw that the super-energy tensors of the various rank Killing-Yano tensors are
trace free parts of some St\"{a}ckel-Killing tensors.
If a trace free St\"{a}ckel-Killing tensor of rank two can be written as
\begin{equation}
K_{\mu\nu}=F_{\mu a}F_\nu^{\ a}-\frac14 g_{\mu\nu}F_{ab}F^{ab}
\end{equation}
i.e. it is required to be the super-energy tensor of some $2$-form $F$,
the starting $2$-form $F$ must satisfies
\begin{eqnarray}\label{condpef}
&&2F_\lambda^{\ a}F_{a(\mu;\nu)}+
2F_\mu^{\ a}F_{a(\nu;\lambda)}+
2F_\nu^{\ a}F_{a(\mu;\lambda)}+\nonumber\\
&&g_{\nu\lambda}F_{ab;\mu}F^{ab}+
g_{\mu\lambda}F_{ab;\nu}F^{ab}+
g_{\mu\nu}F_{ab;\lambda}F^{ab}
=0\;.
\end{eqnarray}
In a more compact form eq. (\ref{condpef}) reads:
\begin{equation}
2F_{a\{(\mu;\nu)}F_{\lambda\}}^{\ \ a}+
F_{ab;\{\mu}g_{\nu\lambda\}}F^{ab}=0\;.
\end{equation}
where curly braces specify a summation over the circular permutations
of enclosed indices.

If for a St\"{a}ckel-Killing tensor $K$ is required to be the super-energy tensor of a
$p$-form with $p\geq2$
\begin{equation}
K_{\mu\nu}=
F_{\mu\rho_1\dots\rho_{p-1}}F_\nu^{\ \rho_1\dots\rho_{p-1}}-
\frac14 g_{\mu\nu}F_{\rho_1\dots\rho_p}F^{\rho_1\dots\rho_p}\;,
\end{equation}
eqs. (\ref{condpef}) becomes:
\begin{eqnarray}
&&\pm2F_\nu^{\ \rho_1\dots\rho_{p-1}}F_{\rho_1\dots\rho_{p-1}(\mu;\lambda)}\pm
2F_\mu^{\ \rho_1\dots\rho_{p-1}}F_{\rho_1\dots\rho_{p-1}(\nu;\lambda)}\pm
2F_\lambda^{\ \rho_1\dots\rho_{p-1}}F_{\rho_1\dots\rho_{p-1}(\mu;\nu)}+\nonumber\\
&&g_{\mu\nu}F_{\rho_1\dots\rho_p;\lambda}F^{\rho_1\dots\rho_p}+
g_{\mu\lambda}F_{\rho_1\dots\rho_p;\nu}F^{\rho_1\dots\rho_p}+
g_{\nu\lambda}F_{\rho_1\dots\rho_p;\mu}F^{\rho_1\dots\rho_p}=0\nonumber
\end{eqnarray}
where +(-) sign is for even(odd) rank $F$ tensors.

It is easy to see that the super-energy tensors obtained from covariantly constant
Killing-Yano tensors $(F_{\rho_1\dots\rho_{p-1}\rho_p;\mu}=0)$
are trace free St\"{a}ckel-Killing tensors, not only trace free parts of some
St\"{a}ckel-Killing tensors (which in general are not by themselves St\"{a}ckel-Killing tensors).\\
For St\"{a}ckel-Killing tensors which are the super-energy tensors of some $p$-forms
we have the following properties (which derives from general properties of super-energy tensors).\\
For any causal, future-oriented vectors $\vec u_1,\vec u_2$ we have DSEP (dominant super-energy property)
\begin{equation}\label{contr-with-temp}
T_{\mu\nu}\{Y_{[2]}\}u^\mu_1u^\nu_2\geq0\;,
\end{equation}
when the strict inequality holds provided that $Y_{[2]}$ is not null.

As a particular case for DSEP, super-energy density for Killing-Yano tensor $Y_{[2]}$
can be obtained.
Super-energy density of $Y_{[2]}$ relative to temporal vector $\vec u$, denoted
by $W_Y(\vec u)$, is defined by
\begin{equation}\label{sedens}
W_Y(\vec u)\equiv T_{\mu\nu}\{Y_{[2]}\}u^\mu u^\nu\geq 0
\end{equation}
and is non negative.
Moreover, if $W_Y(\vec u)=0$ for some $\vec u$
than $T_{\mu\nu}\{Y_{[2]}\}=0$ which implies $Y_{\mu\nu}=0$.\\
Eq. (\ref{sedens}) is important in the following context. It is known that
the geodesic motion i.e.
$\frac{d^2x^\mu}{d\lambda^2}+\Gamma^\mu_{\nu\rho}\frac{dx^\nu}{d\lambda}\frac{dx^\rho}{d\lambda}=0$
admits a quadratic first integral
\begin{equation}\label{quad-first-int}
K_{\mu\nu}\dot x^\mu\dot x^\nu=\;constant
\end{equation}
provided that the symmetric tensor $K_{\mu\nu}$ satisfies the St\"{a}ckel-Killing equation
$K_{(\mu\nu;\lambda)}=0$.
Here $\dot x^\mu$ denote the derivatives of coordinate functions in respect with
the affine parameter of geodesic.

It follows from (\ref{sedens}) that if we consider St\"{a}ckel-Killing tensors
which are the super-energy tensors of some $p$-forms than (\ref{quad-first-int}) provide
a nonnegative first integral.
Even if super-energy density have the mathematical properties of an energy density,
it haven't always the physical dimension and signification of a energy density, this fact
depending of the starting tensor.\\
Relative to a orthonormal coordinate basis ${\vec \epsilon_a}$ such that
$\vec \epsilon_0=\vec u$, we have
\begin{equation}
W_Y(\vec \epsilon_0)=T_{00}\{Y_{[2]}\}=\frac12\sum\limits_{\mu,\nu=0}^3|Y_{\mu\nu}|^2
\end{equation}
If the {\em positive metric} $h_{\mu\nu}$ relative to $\vec u$ is introduced as
\begin{equation}
h_{\mu\nu}(\vec u)\equiv g_{\mu\nu}+2u_\mu u_\nu
\end{equation}
then the super-energy density for $Y_{[2]}$ relative to $\vec u$ can be written
\begin{equation}
W_Y(\vec u)=\frac14 Y_{\mu\nu}Y_{\rho\sigma}h^{\mu\rho}h^{\nu\sigma}.
\end{equation}
Since the divergence of $T\{Y_{[2]}\}$ is
\begin{equation}
T^{\ \mu}_{\lambda\ ;\mu}\{Y_{[2]}\}
=
Y_{\lambda a;\mu}Y^{\mu a}+
Y_{\lambda a}Y^{\mu a}_{\ \ ;\mu}-
\frac12\delta_{\lambda}^{\ \mu}Y_{ab;\mu}Y^{ab}
\end{equation}
it follows that if $Y$ is covariantly constant $(Y_{ab;\mu}=0)$ than
\begin{equation}\label{null-div}
T^{\ \mu}_{\lambda\ ;\mu}\{Y_{[2]}\}=0\;.
\end{equation}
Eq. (\ref{null-div}) can't lead by itself to an integral conservation law since
there is no general Gauss law for rank two or higher tensor fields.
Anyway, if the space admits a Killing vector field i.e. a vector field $\xi_\mu$ for which
$\xi_{(\mu;\nu)}=0$ and if (\ref{null-div}) holds, we have the following local conservation law
\begin{equation}
(T^{\mu\nu}\{Y_{[2]}\}\xi_\nu)_{;\mu}=0
\end{equation}
which via Gauss theorem for vector fields can lead to an integral conservation law.
Moreover, if $Tr(T_{\mu\nu}\{Y_{[2]}\})=0$, than a conformal Killing vector
field is sufficient in order to have the above conservation law.
\subsection {Killing-Yano equations}\label{ky-eq-sec}
As $Y$ is a simple $2$-form, i.e. can be written as a wedge product
of two $1$-forms, we have a single Hodge dual:
\begin{equation}\label{dualky}
Y_{\stackrel\star{\mu\nu}}=\frac12\eta_{\alpha\beta\mu\nu}Y^{\alpha\beta}
\end{equation}
which is a Killing-Yano tensor by itself
$(Y_{\stackrel\star{\mu\nu};\lambda}+Y_{\stackrel\star{\mu\lambda};\nu}=0)$
if $Y_{\mu\nu}$ is covariantly constant.\\
Therefore we have two tensors, the initial $Y$ and it's dual $Y^\star$
for constructing electric and magnetic parts of the Killing-Yano tensor $Y$.\\
The electric part of $Y$ relative to a temporal vector $\vec u$ is defined as
\begin{equation}\label{electricky}
\left(
{}^Y_{\vec u}E
\right)_\nu
\equiv
Y_{\mu\nu}u^\mu.
\end{equation}
If $\vec u$ is covariantly constant,
the electric part ${}^Y_{\vec u}\vec E$ is a Killing vector field $(\left({}^Y_{\vec u}E\right)_{(\mu;\nu)}=0)$.
However, this condition is only a sufficient one.
The necessary condition for the electric part of $Y$ to be a Killing vector field reads
\begin{equation}
Y_{\alpha\mu}u_{;\nu}^\alpha+Y_{\alpha\nu}u_{;\mu}^\alpha=0\;.
\end{equation}
The magnetic part of $Y$ relative to the temporal vector $\vec u$ is
\begin{equation}\label{magneticky}
\left(
{}^Y_{\vec u}H
\right)_\nu
=Y_{\stackrel\star{\mu\nu}}u^\mu
=\frac12\eta_{\alpha\beta\mu\nu}Y^{\alpha\beta}u^\mu\;.
\end{equation}
Since the dual $Y_{\stackrel\star{\mu\nu}}$ of a covariantly constant
Kiling-Yano tensor $Y_{\mu\nu}$ is by itself a Killing-Yano tensor
and the magnetic part of $Y_{\mu\nu}$
is $\left({}^Y_{\vec u}H\right)_\nu=Y_{\stackrel\star{\mu\nu}}u^\mu$
it follows that for a covariantly constant Killing-Yano tensor, the magnetic
part is a Killing vector field in same conditions the electric part is, we just having to
replace $Y_{\mu\nu}$ by $Y_{\stackrel\star{\mu\nu}}$
\begin{equation}
Y_{\stackrel\star{\alpha\mu}}u_{;\nu}^\alpha+Y_{\stackrel\star{\alpha\nu}}u_{;\mu}^\alpha=0\;.
\end{equation}
Both the electric and magnetic part of $Y$ are 4-vectors.
Apparently they have together 8 independent components.
Anyway, since each of them is orthogonal on the chosen
temporal vector field $\vec u$, the number of independent
components is reduced to 6, as is necessary to univocally determine an antisymmetric
$4\times4$ tensor, as Killing-Yano tensor is.\\
Taking into account (\ref{ky-from-eh}) and that
$\eta_{\mu\nu\lambda\rho}=\sqrt{-g}\Delta_{\mu\nu\lambda\rho}$ implies $\eta_{\mu\nu\lambda\rho;\alpha}=0$,
where $\Delta_{\mu\nu\lambda\rho}$ is the Levi-Civita pseudotensor,
the Killing-Yano equation (\ref{defky}) can be written in terms
of $Y$'s electric and magnetic parts as
\begin{eqnarray}\label{ky-eq-by-eh}
&&E_\mu u_{(\mu;\alpha)} - u_\mu E_{(\nu;\alpha)}+E_{\mu;(\nu}u_{\alpha)}-u_{\mu;(\nu}E_{\alpha)}+\nonumber\\
&&\frac12\eta_{\lambda\rho\mu(\nu}u^\lambda_{;\alpha)}H^\rho+
\frac12\eta_{\lambda\rho\mu(\nu}H_{;\alpha)}^\rho u^\lambda=0\;.
\end{eqnarray}
These equations have to be considered together with the orthogonality conditions
between electric, respective magnetic parts of $Y$ and the vector field
$\vec u$, namely
\begin{eqnarray}\label{ortog-cond}
{}_{\vec u}^Y E_\lambda u^\lambda=0,~~~{}_{\vec u}^Y H_\lambda u^\lambda=0\;.
\end{eqnarray}
There is no need to impose the antisymmetry condition $Y_{\mu\nu}=-Y_{\nu\mu}$
in terms of ${}_{\vec u}^Y E$ and ${}_{\vec u}^Y H$, since the expression of
$Y$ from (\ref{ky-from-eh}) is by construction an antisymmetric quantity.

If the electric ${}^Y_{\vec u}E$ and magnetic ${}^Y_{\vec u}H$
parts of a Killing-Yano tensor are determined, the original tensor can be reconstructed
\cite{Senovilla:2000} through
\begin{equation}\label{ky-from-eh}
Y_{\mu\nu}=-2u_{[\mu}E_{\nu]}+
\frac{1}{2}\eta_{\lambda\rho\mu\nu}u^\lambda H^\rho\;.
\end{equation}
If $\vec u$ would be chosen such that $u^\lambda_{;\mu}=0$,
eqs. (\ref{ky-eq-by-eh}) would have a more simplified form.
Such a choice is not always possible since when the metric is required to be a solution
of Einstein equations for empty space, a covariantly constant vector field
is a null type one, or the metric is flat, or in the definitions for
electric and magnetic parts a temporal vector field in required.
Also, the equations would become simpler if $\vec u$ is chosen a to be a Killing vector field.
If $\vec u$ is covariantly constant, the Killing-Yano equations (\ref{ky-eq-by-eh}) becomes
\begin{equation}
E_{\mu;(\alpha}u_{\nu)}-
u_\mu E_{(\nu;\alpha)}+
\frac{\sqrt{-g}}{2}\Delta_{\lambda\rho\mu(\nu}  H_{;\alpha)}^\rho u^\lambda = 0
\end{equation}
which have to be considered together with (\ref{ortog-cond}).

If the Killing-Yano tensor $Y$ is selfdual (anti-selfdual)
it's electric ${}_{\vec u}^Y E$ and magnetic ${}_{\vec u}^Y H$ parts becomes equal
(equal up to a sign) leading to the following Killing-Yano equations
\begin{eqnarray}\label{ky-eq-by-eh-sd}
&&E_\mu u_{(\mu;\alpha)} - u_\mu E_{(\nu;\alpha)}+E_{\mu;(\nu}u_{\alpha)}-u_{\mu;(\nu}E_{\alpha)}\pm\nonumber\\
&&\frac12\eta_{\lambda\rho\mu(\nu}u^\lambda_{;\alpha)}E^\rho\pm
\frac12\eta_{\lambda\rho\mu(\nu}E_{;\alpha)}^\rho u^\lambda=0\;,
\end{eqnarray}
where $+(-)$ sign is for the selfdual (anti-selfdual) case.
\section{Super-energy tensor for a pair of simple $2$-forms}\label{setensor-pair-sec}
Be $A$ and $B$ two simple $2$-forms ($A_{(ab)}=A_{ab}$ respective $B_{(ab)}=B_{ab}$).
We note
\begin{equation}
X\equiv A+B,\ \ \ \ \ Y\equiv A-B.
\end{equation}
It is manifest that $X$ and $Y$ are also simple $2$-forms, so that
the corresponding super-energy tensors can be immediately written:
\begin{eqnarray}
T_{\mu\nu}\{X_{[2]}\}
=&&
X_{\mu a}X^{\ a}_\nu-
\frac14 g_{\mu\nu}X_{ab}X^{ab}=\nonumber\\
&&A_{\mu a}A^{\ a}_\nu+
A_{\mu a}B^{\ a}_\nu+
B_{\mu a}A^{\ a}_\nu+
B_{\mu a}B^{\ a}_\nu-\nonumber\\
&&\frac14g_{\mu\nu}
[
A_{ab}A^{ab}+
A_{ab}B^{ab}+
B_{ab}A^{ab}+
B_{ab}B^{ab}
]=\nonumber\\
&&T_{\mu\nu}\{A_{[2]}\}+T_{\mu\nu}\{B_{[2]}\}+\nonumber\\
&&A_{\mu a}B^{\ a}_\nu+B_{\mu a}A^{\ a}_\nu-
\frac14g_{\mu\nu}A_{ab}B^{ab}-
\frac14g_{\mu\nu}B_{ab}A^{ab}\;.
\end{eqnarray}
Analogously,
\begin{eqnarray}
T_{\mu\nu}\{Y_{[2]}\}=&&T_{\mu\nu}\{A_{[2]}\}+T_{\mu\nu}\{B_{[2]}\}-\nonumber\\
&&A_{\mu a}B^{\ a}_\nu-
B_{\mu a}A^{\ a}_\nu+\nonumber\\
&&\frac14g_{\mu\nu}A_{ab}B^{ab}+
\frac14g_{\mu\nu}B_{ab}A^{ab}\;.
\end{eqnarray}
If we take as definition for the super-energy tensor of a pair of two simple $2$-forms
$A$ and $B$
\begin{equation}
T_{\mu\nu}\{A_{[2]},B_{[2]}\}\equiv
\frac14
\left[
T_{\mu\nu}\{X_{[2]}\}-T_{\mu\nu}\{Y_{[2]}\}
\right]\;,
\end{equation}
we obtain
\begin{equation}\label{setenspt2forme}
T_{\mu\nu}\{A_{[2]},B_{[2]}\}
=
\frac12
\left[
A_{\mu a}B^{\ a}_\nu+B_{\mu a}A^{\ a}_\nu
\right]-
\frac14g_{\mu\nu}A_{ab}B^{ab}\;.
\end{equation}
If the two $2$-forms $A$ and $B$ are taken to be equal,
than (\ref{setenspt2forme})
resemble the standard way of constructing the super-energy tensor of a simple
$2$-form.\\
Since every term in r.h.s. of (\ref{setenspt2forme}) contains
elements from each tensor in pair, it is obvious that (\ref{setenspt2forme})
is a divergence free St\"{a}ckel-Killing tensor if both tensors
in pair are covariantly constant.\\
The above construction can be applied to a pair of two arbitrarily
chosen $p$-forms, the only difference being on summation indices,
namely: whenever we have a contraction of type $A_{\mu a}B_\nu^{\
a}$ we'll replace it with $A_{\mu\rho_2\dots\rho_p}B_\nu^{\
\rho_2\dots\rho_p}$ and instead of
$A_{ab}B^{ab}$ we'll write $A_{\rho_1\dots\rho_p}B^{\rho_1\dots\rho_p}$.\\
The same technique can be used to construct a super-energy tensor for a
pair of two $r$-fold forms which have the same structure of indices blocks, i.e.
both have the same number of blocks and the same number of indices in corresponding
blocks. Anyway, the formulas are rather complicated and we don't give them here.

Since from (\ref{se-tens-with-dual}) the following alternative form of
the super-energy tensor (\ref{setenspt2forme}) can be derived
\begin{equation}\label{setenspt2formedual}
T_{\mu\nu}\{A_{[2]},B_{[2]}\}=
A_{\mu a}B_\nu^{\ a}+B_{\mu a}A_\nu^{\ a}+
A_{\mu a}^{\star}B_{\;\nu}^{\star \;a}+B_{\mu a}^{\star}A_{\;\nu}^{\star \;a}\;,
\end{equation}
it follows that if the starting $2$-forms are simultaneously selfdual or anti-selfdual
the super-energy tensor becomes
\begin{equation}\label{setenspt2formedual1}
T_{\mu\nu}\{A_{[2]},B_{[2]}\}=
2\left[A_{\mu a}B_\nu^{\ a}+B_{\mu a}A_\nu^{\ a}\right]\;,
\end{equation}
while in case when they are one selfdual and the other anti-selfdual the
super-energy tensor vanishes.
\subsection{Application for Euclidean Taub-NUT space}
In a special choice of coordinates the Euclidean Taub-NUT
line element takes the form
\begin{equation}\label{taub-nut-metric}
ds^2 = V(r)\left(dr^2+r^2 d\theta^2+r^2\sin^2\theta d\varphi^2\right)+
16m^2 V^{-1}(r)\left(d\chi+\cos\theta d\varphi\right)^2
\end{equation}
with $V(r)=1 + \frac{4m}{r}$.

There are four Killing vectors for metric (\ref{taub-nut-metric})
\begin{equation}\label{k-vect}
D^{(A)}=R^{(A)\mu}\,\partial_\mu,~~~~A=0,1,2,3,
\end{equation}
where
\begin{eqnarray}
D^{(0)}&=&\partial_\chi\nonumber\\
D^{(1)}&=&-\sin\varphi\,\partial_\theta-
\cos\varphi\,\cot\theta\,\partial_\varphi+
\frac{\cos\varphi}{\sin\theta}\,\partial_\chi\nonumber\\
D^{(2)}&=&\cos\varphi\,\partial_\theta-
\sin\varphi\,\cot\theta\,\partial_\varphi+
\frac{\sin\varphi}{\sin\theta}\,\partial_\chi\nonumber\\
D^{(3)}&=&\partial_\varphi\;.
\end{eqnarray}
The first vector $D^{(0)}$ generates the $U(1)$ of $\chi$ translations and commutes
with the other Killing vectors, while the remaining three vectors,
corresponding to the invariance of the metric (\ref{taub-nut-metric}) under spatial
rotations, obey an $SU(2)$ algebra.
These symmetries would correspond to conservation of the so called
{\em relative electric charge} and the angular momentum:
\begin{equation}
q=16m^2\,V(r)\,(\dot\chi+\cos\theta\,\dot\varphi)
\end{equation}
\begin{equation}
\vec{j}=\vec{r}\times\vec{p}\,+\,q\,{\vec{r}\over r}\;,
\end{equation}
where $\vec{p}=V^{-1}(r)\dot{\vec{r}}$ is the
{\em mechanical momentum} which is only part of the momentum
canonically conjugate to $\vec{r}$.

Taub-NUT space admits the following Killing-Yano tensors
\begin{eqnarray}
f^{(i)}&=&8m\left(d\chi + \cos\theta d\varphi\right)\wedge dx_i -
\epsilon_{ijk}\left(1+\frac{4m}{r}\right) dx_j \wedge dx_k~~~i,j,k=1,2,3\\
f^{(y)}&=&8m\left(d\chi + \cos\theta  d\varphi\right)\wedge dr +
4r\left(r+2m\right)
\left(1+\frac{r}{4m}\right)\sin\theta\;d\theta\wedge d\varphi\;,
\end{eqnarray}
where the first three are covariantly constant, while the last one
has only one non-vanishing component of the field strength
\begin{equation}
f^{(y)}_{r\theta;\varphi} = 2\left(1+\frac{r}{4m}\right)r\sin\theta\;.
\end{equation}

For the Taub-NUT metric there is a conserved vector, analogous to the
Runge-Lenz vector of the Kepler type problem
\begin{equation}\label{runge}
\vec K= \frac1 2 {\vec K}_{\mu\nu} \Pi^\mu \Pi^\nu= \vec p \times
\vec j + \left( \frac{q^2}{4m}-4mE\right)\frac{\vec r}{r}\;,
\end{equation}
where the conserved energy is
\begin{equation}
E=\frac1 2 g^{\mu\nu} \Pi_\mu \Pi_\nu=\frac1 2 V^{-1}(r)\left[
{\dot{\vec r}}^{~2} + \left(\frac{q}{4m}\right)^2 \right].
\end{equation}

Taking $A=f^{(y)}$ and $B=f^{(i)}$ in (\ref{setenspt2forme})
and observing that the last term, being proportional with
metric tensor, leads to trivial constants of motion and can be dropped out,
we obtain the following three nontrivial St\"{a}ckel-Killing tensors
\begin{equation}\label{setens-for-yi}
\tilde T^{(i)}_{\mu\nu}\{f^{(y)},f^{(i)}\}=
\frac12
\left[
f^{(y)}_{\ \ \mu a}f^{(i)a}_{\ \ \nu}+f^{(i)}_{\ \ \mu a}f^{(y)a}_{\ \ \nu}
\right]~~~i=1,2,3\;.
\end{equation}
An explicit evaluation \cite{MVisinescuDVaman:PhysRev:1998} shows that the components of the Runge-Lenz
vector (\ref{runge}) can be written in terms of the geometrical objects
(\ref{setens-for-yi}), (\ref{k-vect})
\begin{equation}
K_{\mu\nu}^{(i)} = 2m\tilde T^{(i)}_{\mu\nu}\{f^{(y)},f^{(i)}\} +
\frac{1}{8m}
\left(R_{\mu}^{(0)} R_{\nu}^{(i)} + R_{\nu}^{(0)} R_{\mu}^{(i)}\right)\;.
\end{equation}
\section{Conclusions}\label{conclusions-sec}
In this paper we have investigated a class of
basic super-energy tensors, namely those constructed from
Killing-Yano tensors. We found that they are trace free parts of
some rank two St\"{a}ckel-Killing tensors and determined the conditions
for these super-energy tensors to be trace free St\"{a}ckel-Killing tensors not
only trace free part of some St\"{a}ckel-Killing tensors.
We have also investigated the special cases when the Killing-Yano tensors
are covariantly constant, self-dual and antiself-dual.
An alternative form for Killing-Yano equations in terms of electric and magnetic
parts of the Killing-Yano tensor was written.
We have also introduced a generalized definition of super-energy tensors
for cases when the starting object is not a single tensor, but a pair of tensors,
which we used to obtain an analogous of the conserved Runge-Lenz
vector from Keppler problem in an Euclidean Taub-NUT space.
\section*{Acknowledgments}
The authors are grateful to Mihai Visinescu for valuable suggestions and discussions.
This work is supported by M.E.C, AEROSPATIAL - NESSPIAC.

\end{document}